\documentclass[jkps,fleqn]{revtex4}
\usepackage[pdftex]{graphicx}
\usepackage{subcaption}
\usepackage{amssymb}
\usepackage{amsmath}
\usepackage{bm}
\usepackage{enumitem}
\usepackage{etoolbox}
\usepackage{caption}

\begin{document}
\setcounter{page}{1}
\title[]{Next-to-leading order QCD predictions for exclusive $J/\psi$ photoproduction in ultraperipheral Pb+Pb collisions at the LHC\footnote{\scriptsize{Presented at DIS2022: XXIX International Workshop on Deep-Inelastic Scattering and Related Subjects, Santiago de Compostela, Spain, May 2-6 2022. Speaker: C. Flett.}}}
\author{Kari J. Eskola$^{1,2}$}
\author{Chris A. Flett$^{1,2}$}
\author{Vadim Guzey}
\author{Topi Löytäinen$^{1,2}$}
\author{Hannu Paukkunen$^{1,2}$}
\address{$^1$University of Jyv\"{a}skyl\"{a}, Department of Physics,
P.O. Box 35, FI-40014 University of Jyv\"{a}skyl\"{a}, Finland}
\address{$^2$Helsinki Institute of Physics, P.O. Box 64, FI-00014 University of Helsinki, Finland}

\begin{abstract}
We present the first study of rapidity-differential cross sections of coherent exclusive $J/\psi$ photoproduction in heavy-ion ultraperipheral collisions (UPCs) at the LHC, $\text{d}\sigma(\text{Pb}+\text{Pb}\rightarrow \text{Pb}+J/\psi+\text{Pb})/\text{d}y$, to NLO in collinear factorisation.  Taking the generalised parton distributions as their forward limit counterparts, the NLO contributions in the cross sections are quantified and we show that the real part of the amplitude and quark-PDF contributions must not be neglected. We further evaluate the uncertainties arising from the scale-choice at both LO and NLO, and compare our results with exclusive $J/\psi$ photoproduction data in Pb+Pb UPCs from LHC. The scale dependence in $\text{d}\sigma(\text{Pb}+\text{Pb}\rightarrow \text{Pb}+J/\psi+\text{Pb})/\text{d}y$ is sizable, but we can find a scale-choice at NLO that reproduces the Pb+Pb UPC data at both Run 1 and Run 2 collision energies.  This process has traditionally been suggested to be a direct probe of nuclear gluon distributions. We show that the situation changes rather dramatically from LO to NLO, where the interplay of both gluon and quark contributions, and in particular their relative signs, play a significant role. We end with some concluding remarks about the outlook of this observable within a tamed collinear factorisation.

\end{abstract}

\pacs{}

\keywords{}

\maketitle

\section{INTRODUCTION}

The exclusive production of heavy vector mesons has long been an attractive observable of study, primarily due to its sharp sensitivity to the gluon parton distribution function (PDF) in the proton. In 1993, it was shown in\cite{Ryskin:1992ui} that to leading logarithmic accuracy (LLA) in perturbative QCD (pQCD), exclusive $J/\psi$ production via $\gamma+p \rightarrow J/\psi+p$ collisions is directly proportional to the square of the gluon PDF. The LLA is coincident with the leading-order (LO) term in the systematic expansion of $\alpha_s$ in pQCD. 
Measured first in fixed target experiments, then at the HERA collider, now at the LHC, and in the future at an EIC and LHeC, the $\gamma+p \rightarrow J/\psi+p$ process at LO therefore functions as a sensitive probe of the gluon PDF at low momentum fractions and scales.

Later, it was realised that the process $\gamma+{\rm Pb} \rightarrow J/\psi+{\rm Pb}$, accessible at the LHC through ultraperipheral $\text{Pb}+\text{Pb}\rightarrow \text{Pb}+J/\psi+\text{Pb}$ collisions, was analogously a good handle on the {\it nuclear} gluon PDF, see e.g.\cite{Adeluyi:2011rt, Guzey:2013qza}, as well as shedding light on the nuclear shadowing phenomenon. Current descriptions of this observable, differential in the rapidity of the $J/\psi$ meson, include models within the Colour Glass Condensate framework as well as LO pQCD.

In these proceedings, based on our study\cite{Eskola:2022vpi}, we present the exclusive $J/\psi$ photoproduction in ultraperipheral Pb+Pb collisions, $\text{Pb}+\text{Pb}\rightarrow \text{Pb}+J/\psi+\text{Pb}$, to next-to-leading order (NLO) in pQCD. For photoproduction, the heavy charm quark mass, $m_c$, provides the hard scale and warrants the use of pQCD in the description of the underlying $\gamma+p$ and $\gamma+$Pb subprocesses. With the latest measurements from LHC for exclusive $J/\psi$ photoproduction in Pb+Pb collisions\cite{ALICE:2013wjo, ALICE:2021gpt, ALICE:2019tqa, ALICE:2012yye, CMS:2016itn, LHCb:2021bfl} reaching a nucleon-nucleon centre-of-mass energy of $\sqrt{s_{\rm NN}} \sim 5$ TeV and a rapidity of $y \sim 4.5$, the Bjorken variable, $x_B$, extends down to $x_B \approx (M_{J/\psi}/\sqrt{s_{\rm NN}})\,\exp(y) \sim 10^{-5}$, where $M_{J/\psi}$ is the mass of the $J/\psi$. 

\section{Set up and theoretical framework  }

Here we discuss our theoretical framework for the exclusive $J/\psi$ photoproduction, $\text{Pb}+\text{Pb}\rightarrow \text{Pb}+J/\psi+\text{Pb}$, to NLO in pQCD. The initial state Pb nuclei interact in an ultraperipheral collision, driven by the quasi-elastic hard scattering $\gamma+{\rm Pb} \rightarrow J/\psi+{\rm Pb}$ subprocess.  An incoming on-shell photon from either of the Pb nuclei fluctuates into a $c\bar{c}$ heavy quark pair, which interacts with the target Pb nuclei through a colourless two-parton exchange mechanism. In the Bjorken limit at leading twist, following\cite{Ivanov:2004vd}, the probed parton momentum fractions can be expressed as $x+\xi$ and $x-\xi$ so that the net momentum transfer is $\Delta^\mu = 2\xi p^\mu$, where $\xi$ is the skewness parameter and $p^\mu$ is a (lightlike) vector along the collision axis. The Mandelstam variable,  $t = \Delta^2 = 0$.  In the collinear factorisation formula, a given $\xi \approx x_B/2$ entails an integration over $x$ between the quark and gluon NLO coefficient functions and the corresponding generalised parton distribution functions (GPDs). The resulting $c\bar{c} \rightarrow J/\psi$ transition is calculated within LO Non Relativistic QCD (NRQCD), which provides sufficient accuracy\cite{Hoodbhoy:1996zg} once the leading NRQCD colour singlet matrix element $\langle O_1 \rangle_{J/\psi}$ is normalised to the dilepton decay width of the $J/\psi$. 
Due to the lack of experimental Pb tagging, the initiating photon can come from either the left moving Pb or right moving Pb. For a given rapidity, say $y>0$, of the $J/\psi$, there are therefore two configurations that contribute, where the vector meson goes in the direction of the photon ($W_+$) or against it ($W_-$), where $W$ is the $\gamma+p$ centre-of-mass energy. 

The rapidity differential cross section, $\text{d}\sigma/\text{d}y$, is therefore a sum of two terms, 
\begin{equation}
\frac {\text{d}\sigma}{\text{d}y} (\text{Pb}+\text{Pb}\rightarrow \text{Pb}+J/\psi+\text{Pb}) = k^+ \frac{\text{d}N}{\text{d}k^+} \sigma(W_+) + k^- \frac{\text{d}N}{\text{d}k^-} \sigma(W_-),
\end{equation}
where $k^\pm dN/dk^\pm$ are Weizs\"{a}cker-Williams photon fluxes, $k^\pm$ are the photon energies and $\sigma(W_\pm)$ are the hard scattering cross sections. Here, we explicitly perform the integration over the whole impact parameter plane and account for the probability of no additional hadronic interactions or interactions involving the spectator partons inside the initial state Pb nucleus that would otherwise populate the rapidity gap between the produced vector meson and the outgoing Pb and destroy the exclusivity of the event. 

The NLO quark $q$ and gluon $g$ coefficient functions $T_{q,g}$ for the exclusive heavy vector meson photoproduction, used here, are given in\cite{Ivanov:2004vd}. See\cite{Chen:2019uit, Flett:2021ghh} for the corresponding electroproduction ones, where the incoming photon is off-shell and so relevant for $e+p$ collisions at HERA and for the upcoming $e+p$ programme of the Electron Ion collider. The $T_{q,g}$ here are convoluted with the quark singlet and gluon GPDs, $F_{q,g}$, to give the amplitude $\mathcal M$ to NLO,
\begin{equation}
\mathcal M \sim \sqrt{ \langle O_1 \rangle_{J/\psi} } \int_{-1}^1 \text{d}x \left[ T_g(x,\xi,\mu_R,\mu_F) F_g(x,\xi,t,\mu_F) + T_q(x,\xi,\mu_R,\mu_F) F_q(x,\xi,t,\mu_F)\right],
\end{equation}
where $\mu_F$ and $\mu_R$ are the factorisation and renormalisation scales, respectively. 
In this baseline approach, the GPDs are taken in their forward limit $(\xi, t \rightarrow 0)$,
\begin{align}
    F_g(x,0,0,\mu_F) &= F_g(-x,0,0,\mu_F) = xg(x,\mu_F), \nonumber \\
    F_q(x,0,0,\mu_F) &= u(x,\mu_F) + d(x,\mu_F) + s(x,\mu_F) + c(x,\mu_F), \nonumber \\
    F_q(-x,0,0,\mu_F) &= -\bar{u}(x,\mu_F) - \bar{d}(x,\mu_F) - \bar{s}(x,\mu_F) - \bar{c}(x,\mu_F),
\end{align}
where $x>0$ and the corresponding PDFs are taken from the LHAPDF6 interface\cite{Buckley:2014ana}. 

The hard scattering cross sections are finally given as
\begin{equation}
\sigma(W_\pm) = \frac{\text{d} \sigma}{\text{d} t}\biggl|_{t=0}\, \int_{t_{\text{min}}}^\infty \text{d}t\, |F_A|^2,\,\,\,\,\,\,\,\,\,\,\,\,\text{where}\,\,\,\,\,\,\,\,\,\,\,\,\frac{\text{d} \sigma}{\text{d} t}\biggl|_{t=0} = \frac{|\mathcal M|^2}{16 \pi W^4},
\end{equation}
and $F_A$ is the nuclear form factor, calculated as the Fourier transform of a Woods-Saxon distribution. We take $\mu_R = \mu_F$ and check our results with two different methods. The coefficient functions are complicated functions involving the standard $+i\varepsilon$ prescription, which in our numerical implementation are handled as follows.  We take the $\varepsilon$ parameter numerically closer and closer to zero until the results do not change.  This was cross checked in an approach where the complex integrands are series expanded around the branch points $x = \pm \xi$ to show that the singular points within the integration domain are indeed integrable. 



\section{Results}

\begin{figure*}
        \centering
        \begin{subfigure}[b]{0.4\textwidth}
            \centering
            \includegraphics[width=\textwidth]{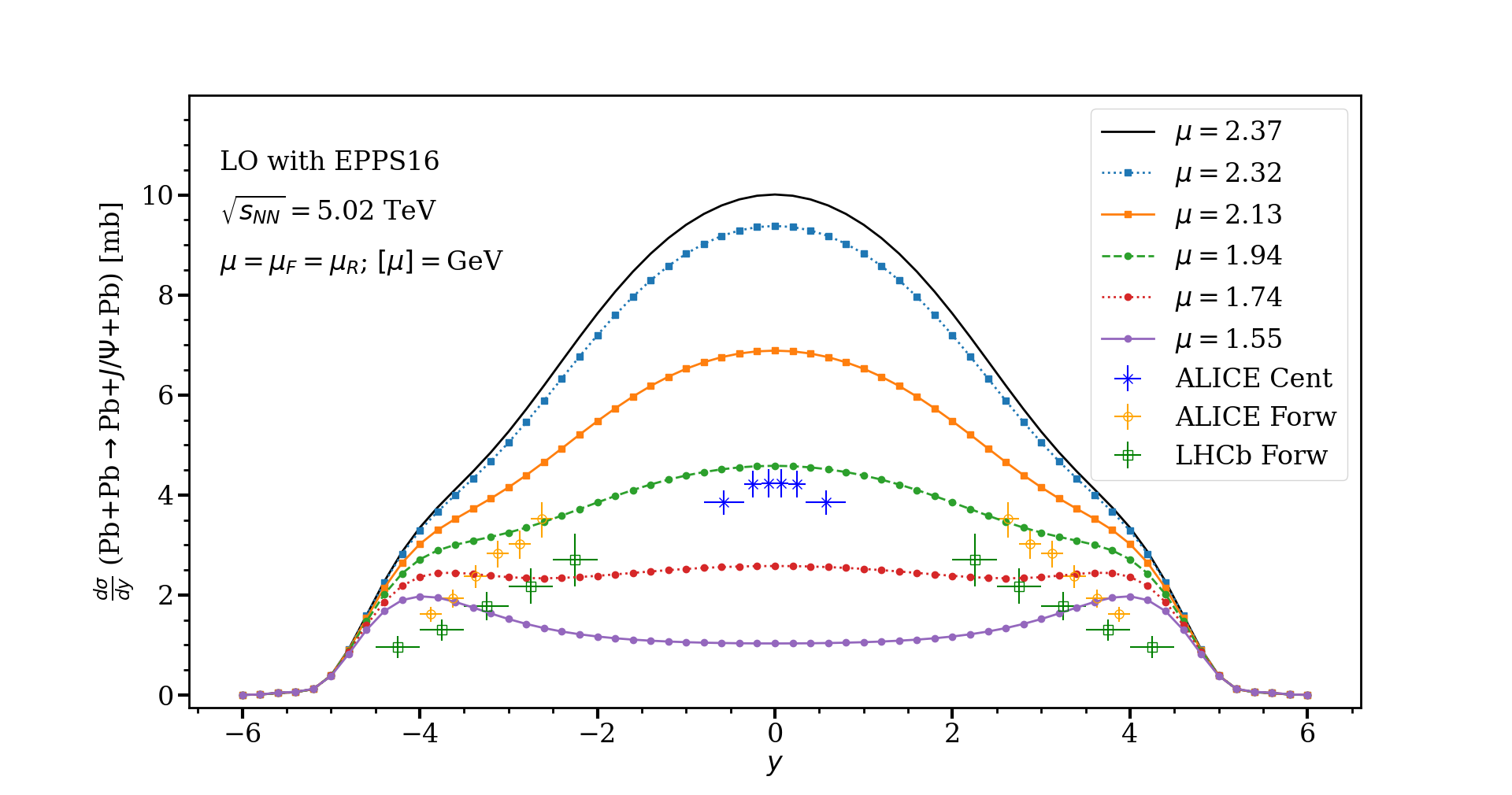}
            {{\small}}    
            \label{fig:jpsi5decomps}
        \end{subfigure}
        \begin{subfigure}[b]{0.4\textwidth}  
             \centering
            \includegraphics[width=\textwidth]{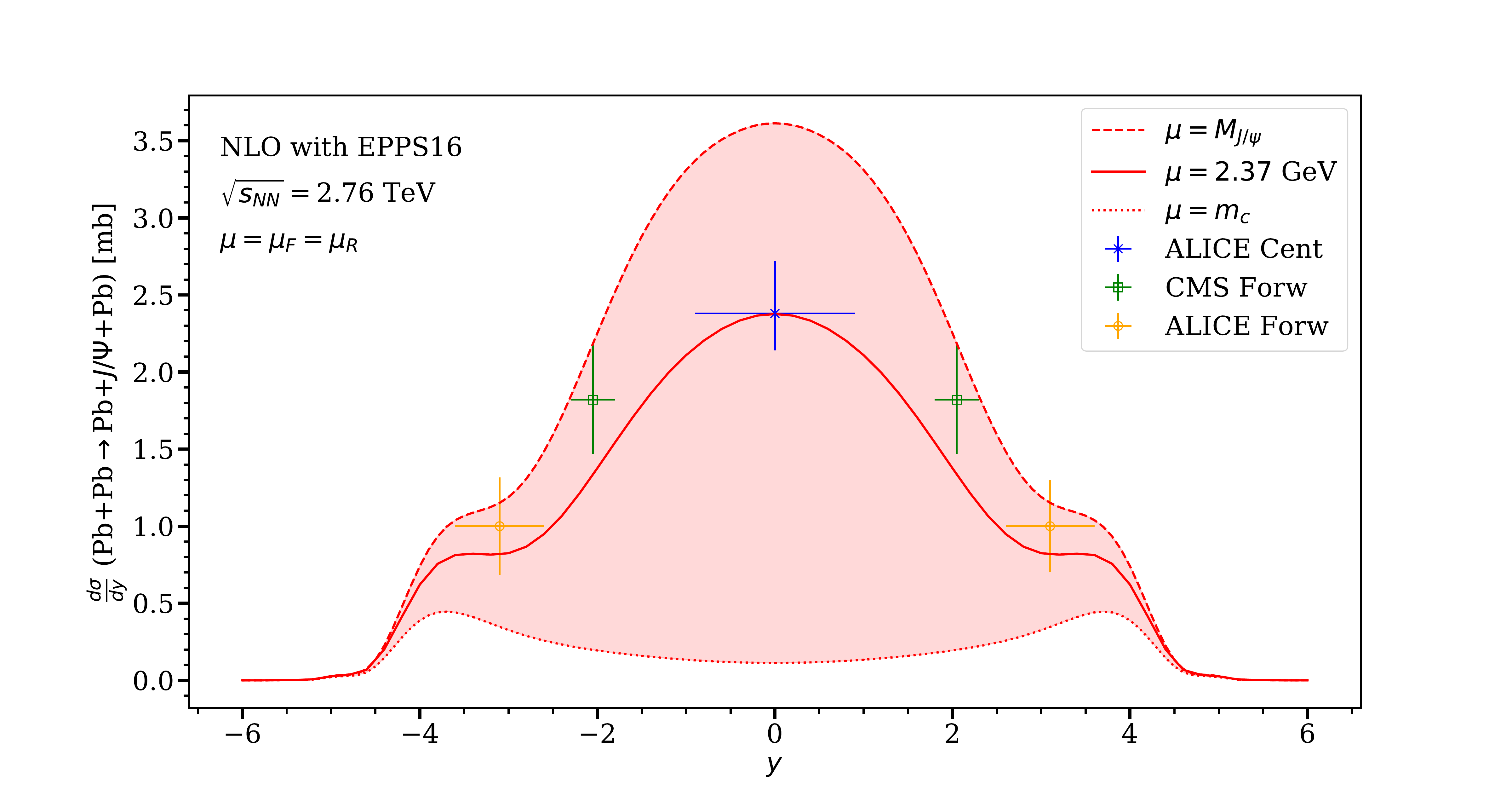}
            {{\small }}    
            \label{fig:jpsi5data}
        \end{subfigure}
        \vskip\baselineskip
        \begin{subfigure}[b]{0.4\textwidth}   
             \centering
            \includegraphics[width=\textwidth]{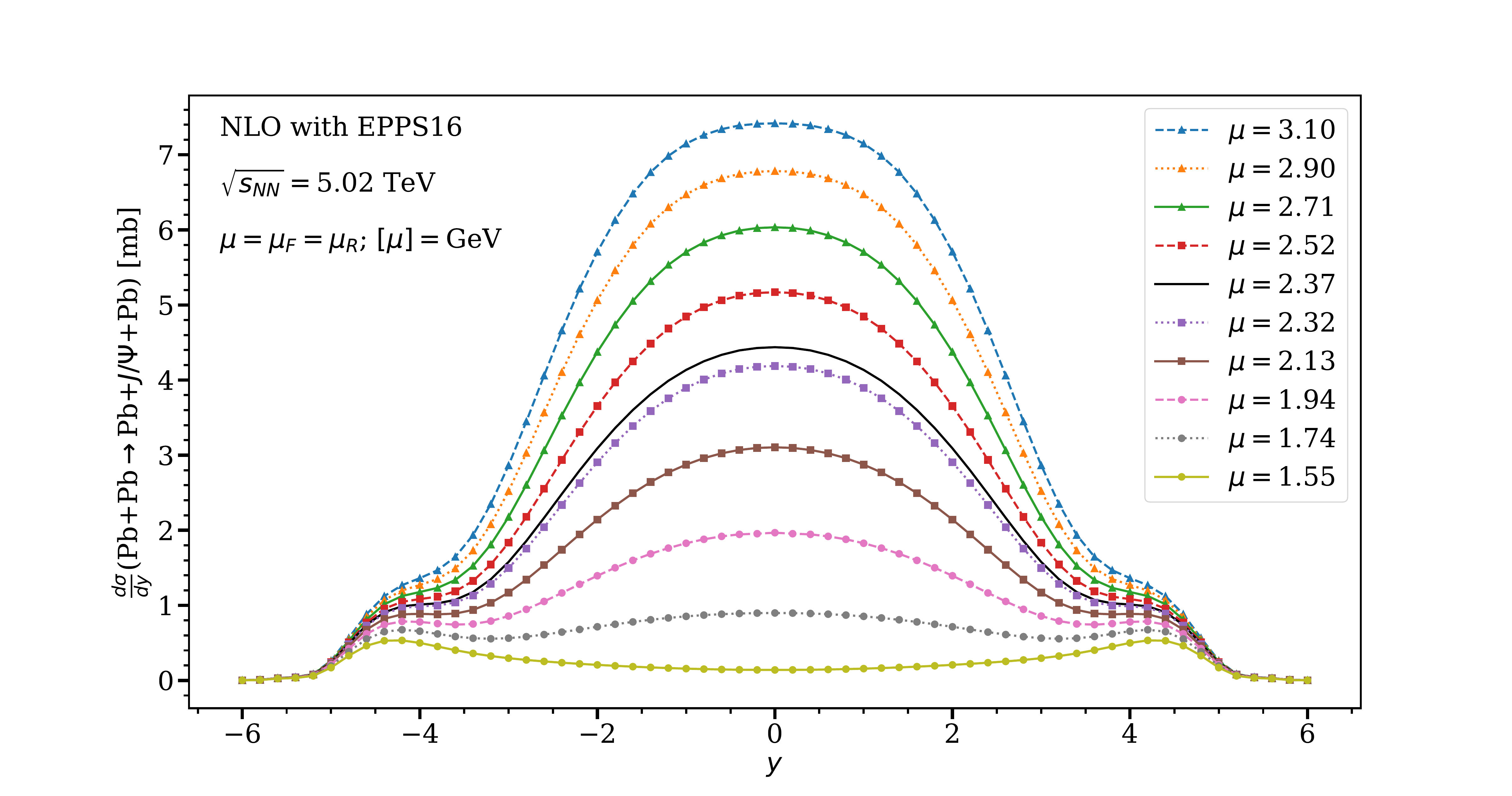}
            {{\small }}    
            \label{fig:ups5decomps}
        \end{subfigure}
        \begin{subfigure}[b]{0.4\textwidth}   
             \centering
            \includegraphics[width=\textwidth]{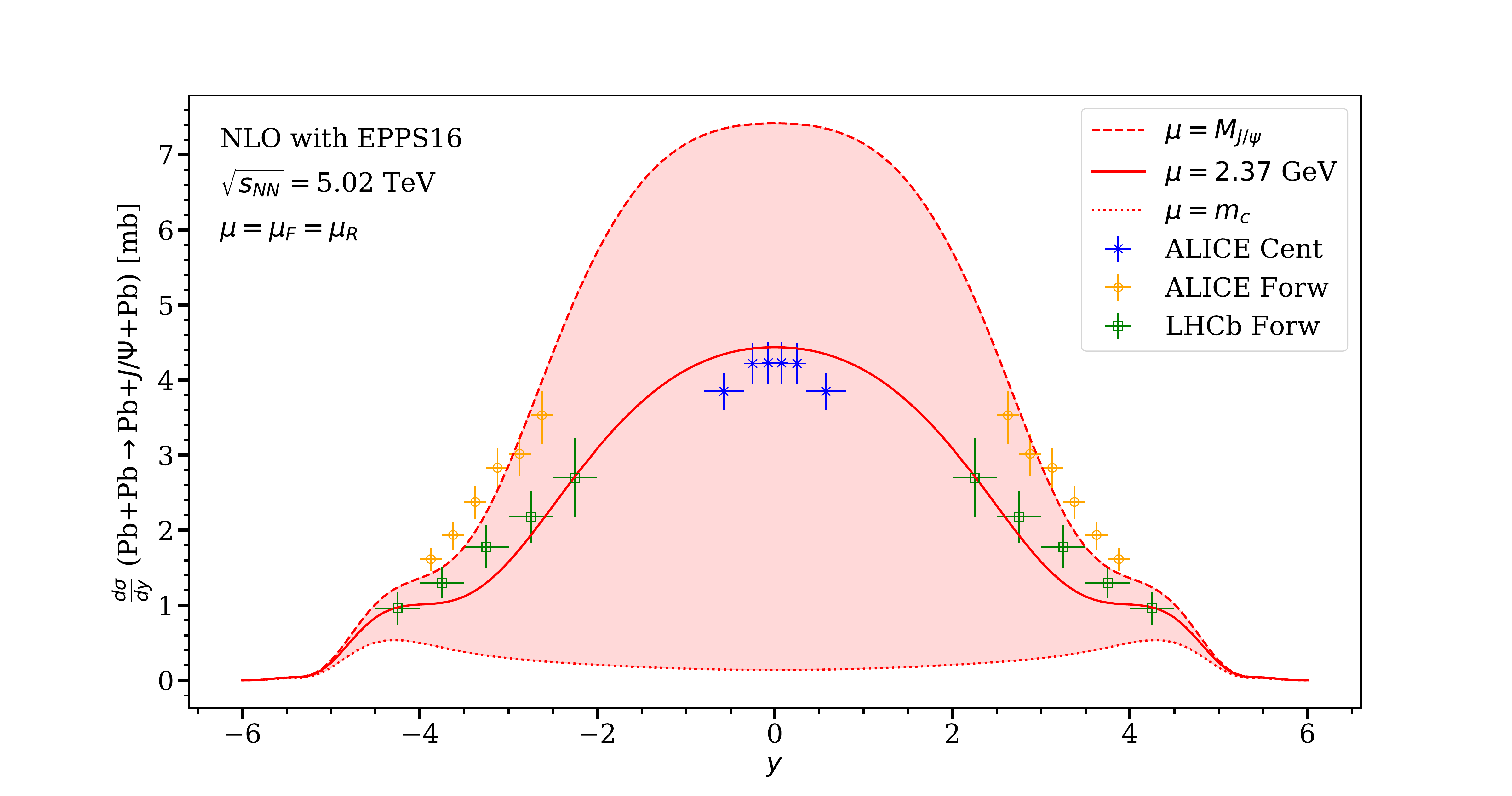}
            {{\small }}    
            \label{fig:ups5data}
        \end{subfigure}
        \caption[  ]
        {Right: Rapidity-differential coherent exclusive $J/\psi$ photoproduction cross section vs. rapidity in Pb+Pb UPCs at $\sqrt{s_{\text{NN}}} = 2.76$ TeV (right, upper) and 5.02 TeV (right, lower), computed with EPPS16 nPDFs, scales $\mu = M_{J/\psi}/2,\,\, 2.37$ and $M_{J/\psi}$. Left: explicit scale variation curves at LO (left, upper) and NLO (left, lower). Note the choice of scales presented at LO for $\sqrt{s_{\text{NN}}} = 5.02$ TeV and therefore the different scale on the y-axis.  } 
        \label{fig:1}
    \end{figure*}

We now proceed with the presentation of the main results of our study. Figure~\ref{fig:1} shows the scale dependence of the rapidity differential cross section, $\text{d}\sigma/\text{d}y$ vs. $y$, at both LO and NLO using the EPPS16 nuclear PDFs\cite{Eskola:2016oht}. The right panel shows the NLO scale uncertainty band resulting from variations of the scale  $\mu = \mu_F = \mu_R$ from $m_c = M_{J/\psi}/2$ up to $M_{J/\psi}$ at both Run 1 (right, upper) and Run 2 (right, lower) collision energies of 2.76 and 5.02 TeV respectively. The scale choice $\mu = \mu_F = \mu_R$ = 2.37 GeV is dubbed ‘optimal’ scale in the following in the sense that at NLO it simultaneously reproduces well the data from ALICE\cite{ALICE:2013wjo, ALICE:2021gpt, ALICE:2019tqa, ALICE:2012yye}, CMS\cite{CMS:2016itn} and LHCb\cite{LHCb:2021bfl} at both collision energies, as shown. 

The left panel shows explicit scale variation curves at LO (left, upper) and NLO (left, lower) at $\sqrt{s_{\text{NN}}} = 5.02$ TeV. The curves are shown over the whole range between $m_c$ and $M_{J/\psi}$ at NLO while at LO it is clear that no curve nicely reproduces the data, and the curve generated using the optimal scale lies consistently above the data. 

The large scale dependence exhibited by this observable was first discussed in\cite{Ivanov:2004vd}, in the context of $\gamma+p$ collisions. The scale-variation in the interval $[m_c, M_{J/\psi}]$ gives rise to a factor of 20 between the predictions in the LO case at mid-rapidity, while in the full NLO result the change is about a factor of 50. This confirms again the conclusions of\cite{Ivanov:2004vd}, that the NLO corrections do not stabilise the results.  It is encouraging, however, that the specific scale choice $\mu = 2.37$ GeV, found through a simultaneous rough fit to the Pb+Pb rapidity differential data at both collision energies, agrees favourably with the experimental distributions. The precise numerical value of this scale obtained is of course susceptible to change with more explicit GPD modelling, but we emphasise here that a scale choice in the vicinity of the mass of the charm quark or $J/\psi$ can be found. 

\begin{figure} [h]
\centering
\includegraphics[scale=0.176]{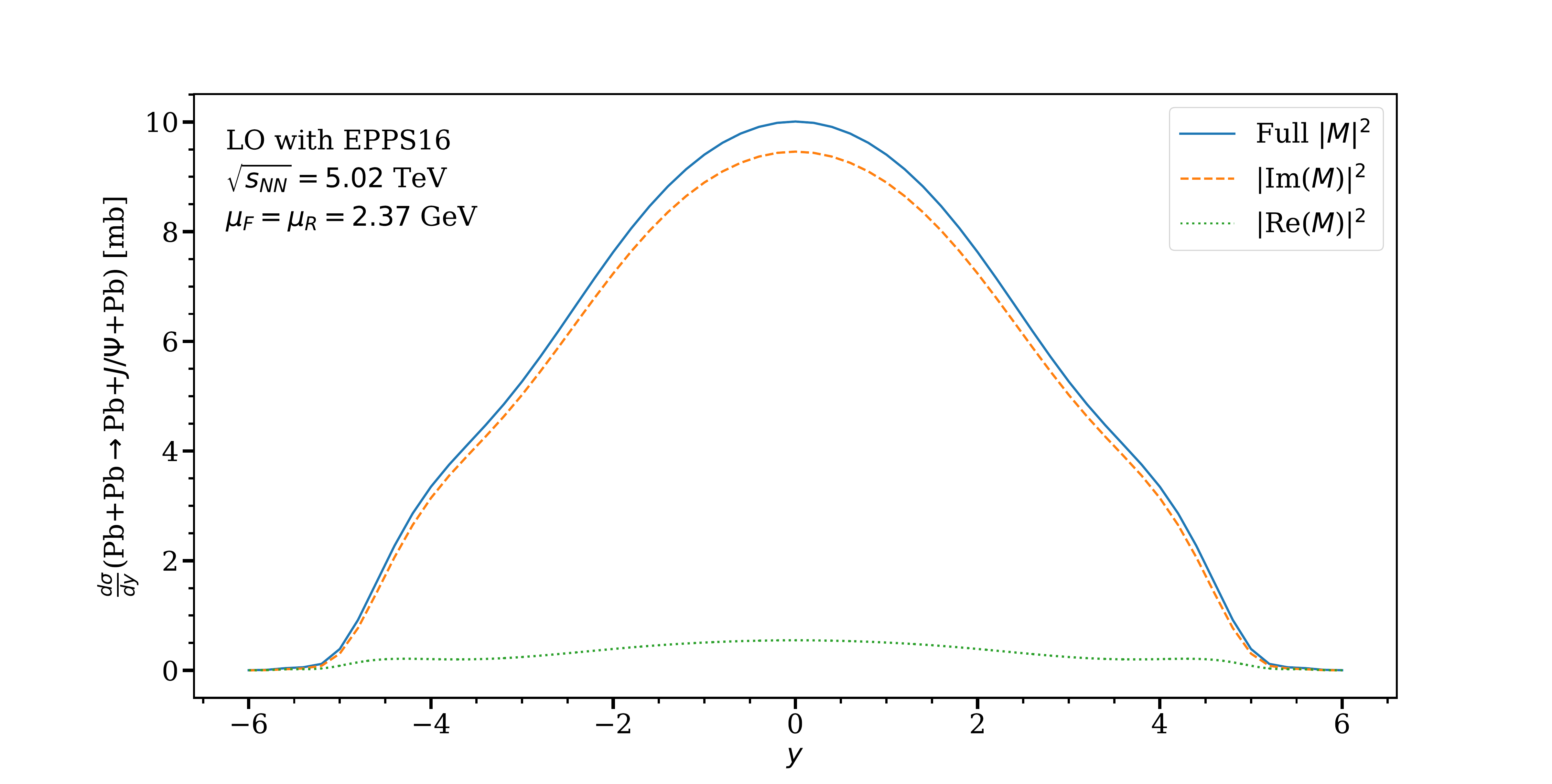}
\qquad
\includegraphics[scale=0.176]{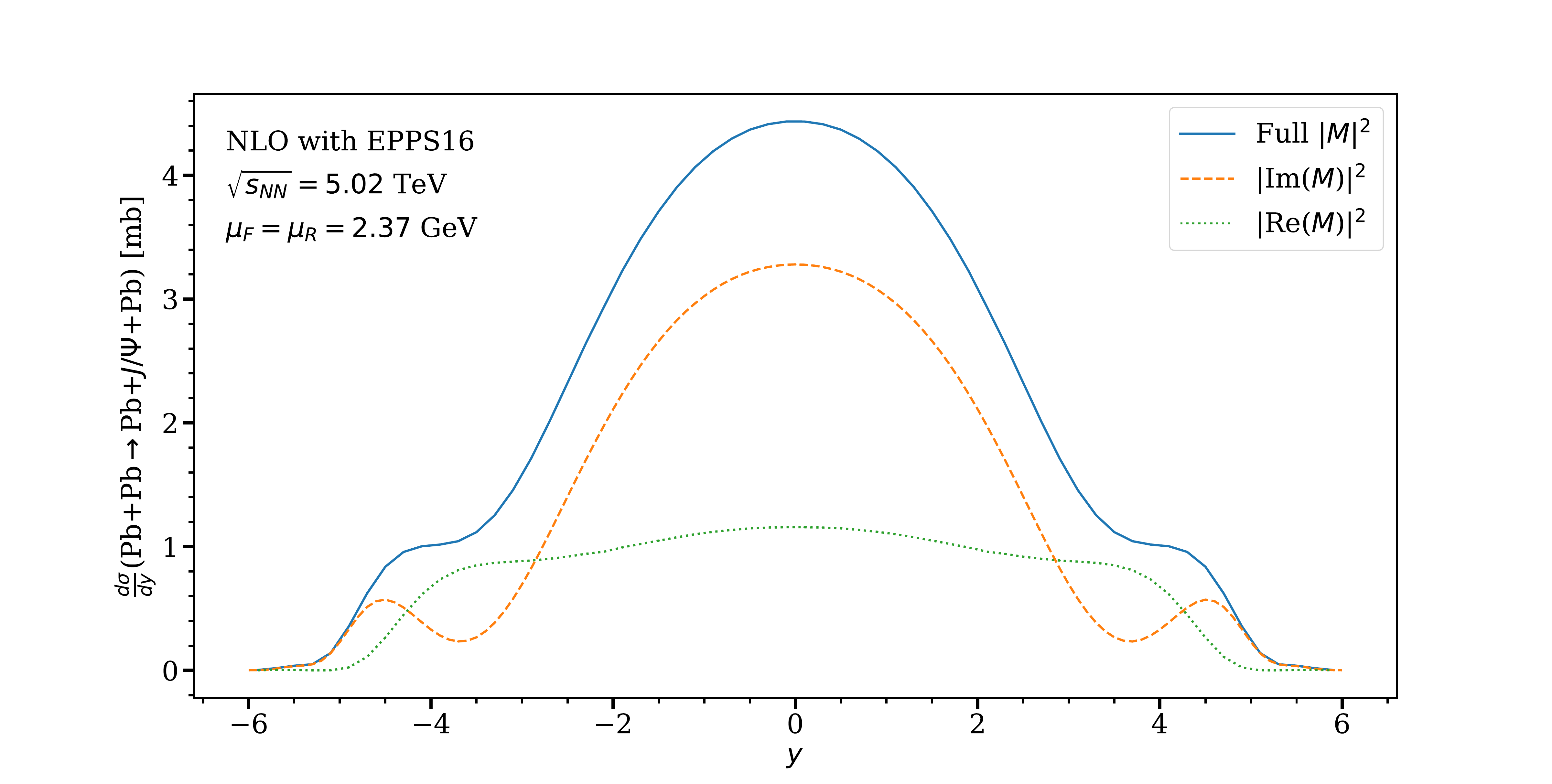}
\caption{Breakdown of the LO (left) and NLO (right) cross sections shown in Fig. 1 into contributions from the imaginary and real parts of the amplitude at the optimal scale.}
\label{fig:2}
\end{figure}

In Fig.~\ref{fig:2}, we show the decomposition of the rapidity differential cross section into real and imaginary parts at both LO (left) and NLO (right) using EPPS16 nPDFs at the optimal scale with $\sqrt{s_{\text{NN}}} = 5.02$ TeV. At LO, it is clear that the imaginary part dominates the real part over the whole range of rapidity considered. At NLO, however, the situation changes dramatically with the real part contributing $\approx 25\%$ at mid-rapidity and there exists shoulder regions around $|y| = 4$ where the real part is larger than the imaginary part. 

\begin{figure} [h!]
\begin{center}
\includegraphics[width=0.6\textwidth]{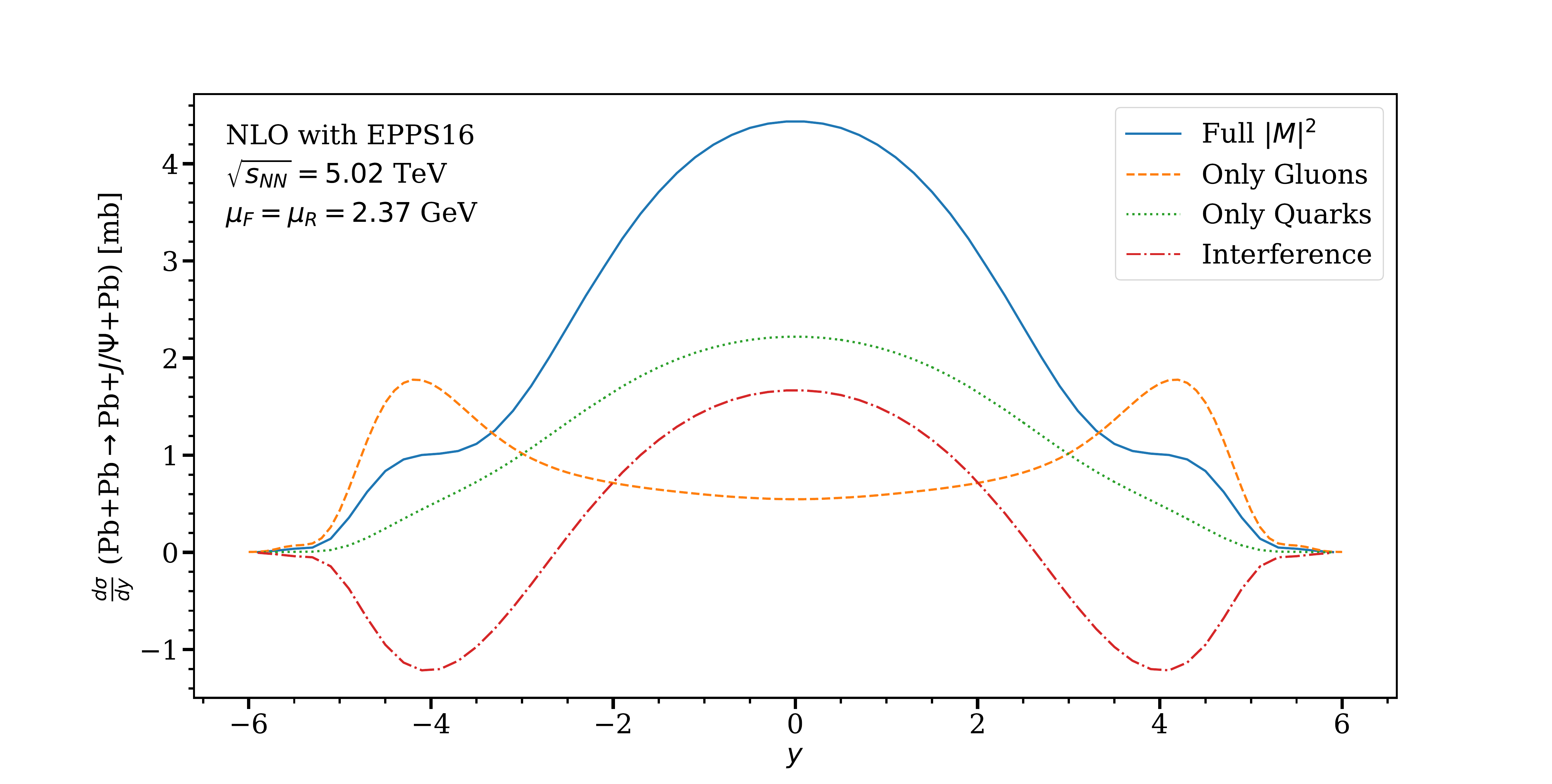}
\caption{Breakdown of the NLO cross section from Fig. 1 into the quark, gluon and quark-gluon interference contributions at the optimal scale. }
\label{fig:3}
\end{center}
\end{figure}

The LO and NLO gluon contributions turn out to have opposite signs and this, as well as indicating a lack of perturbative convergence in $\alpha_s$ for this process, plays a crucial role in Fig.~\ref{fig:3}.
This figure shows how the interplay of the quark and gluon NLO contributions, as well as their interference, produces the total rapidity differential cross section.  What is perhaps particularly striking is that while at LO the process is driven completely by a two-gluon exchange, at NLO when the quark contribution enters the picture, we obtain a complicated mix of contributions over the entire range of rapidity. Most noteworthy is the dominance of the quark contribution around mid-rapidity. This follows from the poor perturbative stability mentioned above and reflects the large extent of cancellation between the LO and NLO gluon amplitudes around $y \approx 0$. We have checked that the LO and NLO gluon amplitudes are larger than the quark amplitudes in absolute terms, but in the construction of the amplitude to NLO it is instead the NLO quark contribution that dominates the combined LO and NLO gluon contributions.  Moreover, the appearance of the shoulder regions at $|y| = 4$ arise from an intricate interplay of the photoproduction cross sections, photon fluxes, nuclear form factor and $W_\pm$ components, see\cite{Eskola:2022vpi} for more details. Interestingly, these shoulders also arise in a description of $\rho$ photoproduction in Pb+Pb UPCs\cite{Frankfurt:2015cwa}. 

\begin{figure} [h]
\centering
\includegraphics[scale=0.176]{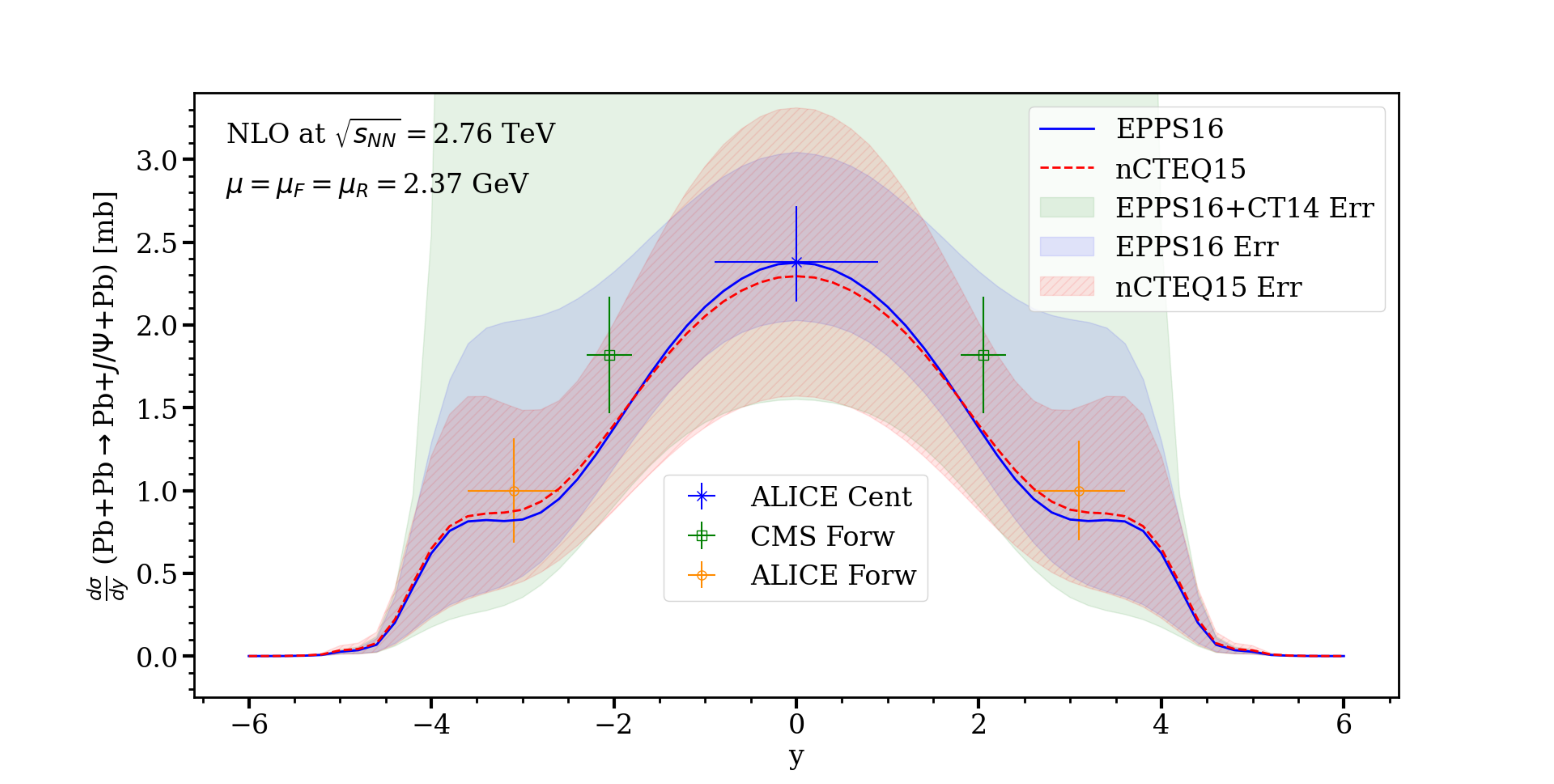}
\qquad
\includegraphics[scale=0.176]{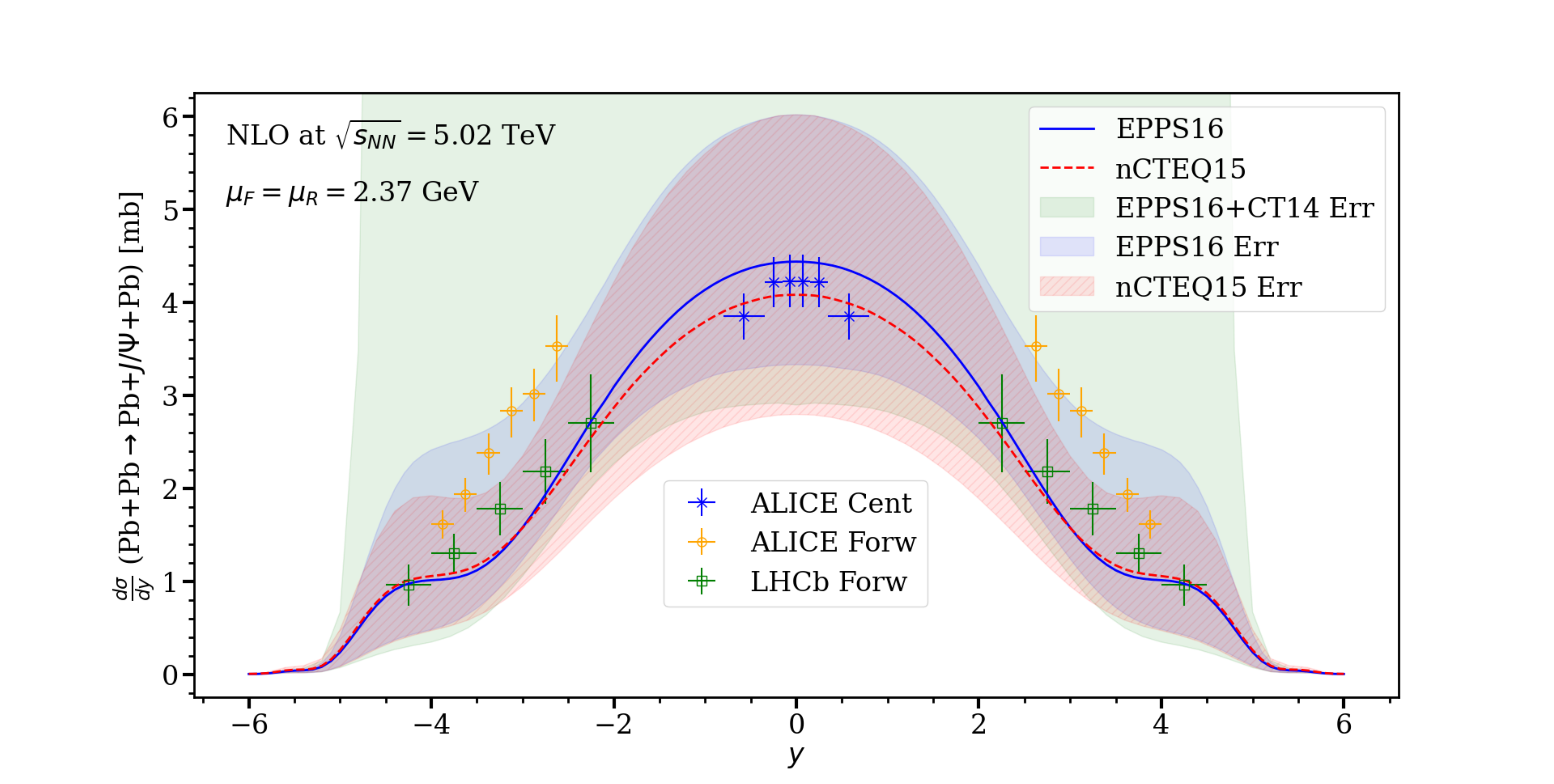}
\caption{Rapidity-differential coherent exclusive $J/\psi$ photoproduction cross section vs. rapidity in Pb+Pb UPCs at $\sqrt{s_{\text{NN}}} = 2.76$ TeV (left) and 5.02 TeV (right) at the optimal scale and with PDF uncertainties included.}
\label{fig:4}
\end{figure}

We finish our presentation of the results by showing the nuclear and total PDF uncertainties alongside our central predictions at Run 1 (left) and Run 2 (right) collision energies, at the optimal scale at NLO, in Fig.~\ref{fig:4}.  The solid and dashed lines correspond to the central predictions generated using the EPPS16\cite{Eskola:2016oht} and nCTEQ15\cite{Kovarik:2015cma} PDFs, respectively. The shaded blue (orange) band gives the EPPS16 (nCTEQ15) nuclear PDF uncertainties and the green band is obtained by including also the free proton CT14NLO PDF\cite{Dulat:2015mca} errors, which as shown, are large and dominated by a single PDF LHAPDF error set. The nuclear PDF uncertainty bands nicely encompass the data at both collision energies but there exists a tension between the ALICE and LHCb forward rapidity data at $\sqrt{s_{\text{NN}}} = 5.02$ TeV.\footnote{\tiny{See talk by X. Wang, for the LHCb collaboration, in this conference, where the update of the LHCb measurement\cite{LHCb:2022ahs} was presented.}}


\section{Conclusions and outlook}

We have presented the first implementation of exclusive $J/\psi$ photoproduction in Pb+Pb collisions to NLO in pQCD.\footnote{\tiny{We acknowledge the financial support from the Magnus Ehrnrooth foundation (T.L.), the Academy of Finland projects 308301 (H.P.) and 330448 (K.J.E.). This research was funded as a part of the Center of Excellence in Quark Matter of the Academy of Finland (project 346325). This research is part of the European Research Council project ERC-2018-ADG-835105 YoctoLHC.}} The large scale dependence exhibited by this observable was emphasised, as well as the dominance of the quark NLO contribution around mid-rapidity. Recalling one of the original motivations for the study of exclusive $J/\psi$ photoproduction, amounting to providing unprecendented constraints on the gluon PDF so that ultimately the experimental data may be used to improve global PDF analyses of the proton and heavy nuclei at low momentum fractions and scales, it appears the above deters this ideology. It should, however, be emphasised that here we have worked within the {\it conventional} collinear factorisation approach, where the notable features of our study are well known and qualitatively reflected in\cite{Ivanov:2004vd} in the context of the underlying $\gamma+p \rightarrow J/\psi+p$ process. A systematic taming of standard collinear factorisation amounting to a programme of low $x$ resummation and implementation of a crucial infrared subtraction, proposed for this process in\cite{Flett:2019pux, Flett:2020duk}, provided a much milder scale dependence and a small quark NLO contribution over the whole kinematic range currently accessible by LHC. As shown in\cite{Flett:2019pux, Flett:2020duk}, this allowed for a fruitful description of the $\gamma+p$ photoproduction data, and then subsequently an extraction of a low $x$ and low scale gluon PDF using the $p+p$ UPC data. This modified collinear factorisation framework was more recently applied to $p$+Pb collisions in\cite{Flett:2021fvo, Flett:2022ues} and, in future works, as well as incorporating explicit GPD modelling, can be carried through to our baseline Pb+Pb study presented here. This can be expected to be desirable for the heavy-ion physics community and provide additional constraints on the nuclear gluon PDF and lead to an improved understanding of the shadowing phenomenon. 






\end{document}